# Modulation of Quantum Transport in Complex Oxide Heterostructures with Proton Implantation

Haidong Liang[1,2,#], Ganesh Ji Omar[1, #], Kun Han[3], Andrew A. Bettiol[1,2,*], Zhen Huang[3,4,*], Ariando[1,*]

[1]Department of Physics, National University of Singapore, Singapore 117551, Singapore

[2]Centre for Ion Beam Applications (CIBA), Department of Physics, National University of Singapore, Singapore 117542, Singapore

[3]Leibniz International Joint Research Center of Materials Sciences of Anhui Province, Institutes of Physical Science and Information Technology, Anhui University, Hefei 230601, China

[4]Stony Brook Institute at Anhui University, Anhui University, Hefei 230039, China

[#] These authors contributed equally to this work.

[*]Contact Author: a.bettiol@nus.edu.sg, huangz@ahu.edu.cn, phyarian@nus.edu.sg

**Abstract:**

The interfacial electronic properties of complex oxides are governed by a delicate balance between charge transfer, lattice distortions, and electronic correlations, posing a key challenge for controlled tunability in materials research. Here, we demonstrate that proton implantation serves as a precise tool for modulating interfacial transport in SrTiO$_3$-based heterostructures. By introducing protons into the SrTiO$_3$ substrate beneath an amorphous (La,Sr)(Al,Ta)O$_3$ capping layer, we uncover a competition between disorder and charge doping induced by implantation. At low implantation fluences below $1\times10^{15}$ protons/cm$^2$ (1E15), charge doping dominates, leading to an increase in carrier density and mobility, analogous to electrostatic gating effect. This enables the emergence of quantum transport oscillations at low temperature. Conversely, at higher fluences (above 1E15), disorder scattering prevails, suppressing carrier mobility and inducing an insulating state. The nonmonotonic evolution of transport with implantation fluence underscores the critical interplay between electronic correlations and disorder, offering a new paradigm for the controlled engineering of interfacial quantum states in SrTiO$_3$-based oxide heterostructures.

**Introduction**

Continuous effort has been made to develop the technique of ion implantation to fabricate commercial semiconductor devices since 1957.[1–3] One example is the widely-used silicon-on-insulator substrate, which can be obtained using oxygen ion implantations followed by a high-temperature annealing.[4] Another example is the smart-cut process, where the high-dose ion implantation is applied to create a cracking layer at a specific location to induce an in-depth splitting in the target sample.[5] Moreover, it has been demonstrated that over 40 steps of ion implantations, with various doses and energies, are required to achieve a modern 28-nm "system on a chip" device.[6] So, ion implantation has played an important role in developing novel functionalities and device fabrications in Si-based industry.[2] On the other hand, the oxide heterointerface is capable of integrating multiple functionalities into one device and has been proposed as a possible solution to preserve Moore's law in future.[7] So, it is curious to clarify whether the ion implantation, a fully-developed technique in modern semiconductor industry, can be applied to functional oxide heterointerfaces for designing the next-generation electronic devices.

A good example of functional oxide heterointerface is the $SrTiO_3$-based interface, where multiple properties including the two-dimensional (2D) conductivity,[8,9] magnetism,[10,11] superconductivity,[12] ferroelectricity[13] and spin-orbital coupling[14–16] are coexisting. There are several reports investigating the ion-implantation effect on the well-known conducting $LaAlO_3/SrTiO_3$ interface. Mathew *et al*. used 2 MeV protons with dose above $6\times10^{17}$ protons/cm$^2$(6E17), or 500 keV He ions with dose above 1E16, to remove the interfacial conductivity of exposed areas.[17] Similarly, Hurand *et al.* applied oxygen ions (50 keV, $5\times10^{12}$ cm$^{-2}$) to pattern the $LaAlO_3/SrTiO_3$ interface for obtaining the top-gated field-effect transistor, of which the micro-size channel protected from the ion implantation maintains the metallic transport behavior.[18] Also, Aurino *et al.* studied the post thermal annealing, which heals the ion-implantation-induced damages to recover the interfacial conductivity.[19,20] All those studies focus on the ion-implantation-induced *structural damage*, which creates disorders for carrier localizations at the ion-implanted SrTiO$_3$-based interface. However, the other side of ion implantations, *charge doping*, at oxide heterointerfaces, is not fully discussed. During the ion implantation, the high-energy ions will knock out the oxygen in oxides, leaving oxygen vacancies (as localized positive charges) and excited electrons (as mobile negative charges) in SrTiO$_3$. It has been well documented that the insulating SrTiO$_3$ can be easily turned into a conductor by various types of electron doping, including chemical substitutions or electrostatic gating.[21] In this work, we will present and discuss about two sides of the ion-implantation effect, *structural damage* and *charge doping*, which simultaneously affect the SrTiO$_3$-based interface.

**Results and discussions**

We used 50 keV protons (or H$^+$ in some figures) for ion implantation, and the target oxide heterointerfaces are prepared by growing the amorphous (La,Sr)(Al,Ta)O$_3$ (*a*-LSAT) layer on the proton-implanted (001) SrTiO$_3$ substrate with different implantation doses. If implantation were performed after deposition, the implanted protons would traverse the already formed conducting interface and severely disrupt it, rendering the interface insulating. **Figure 1(a-c)** summarize the process of sample preparation. First, the SrTiO$_3$ substrate was treated by the buffered HF and thermal annealing to achieve an atomically flat TiO$_2$-terminated surface. Second, the protons were implanted into the treated SrTiO$_3$ substrate with different doses, ranging from 1E14 to 1E16. **Figure 1(d)** presents the gradual change of colors in proton-implanted SrTiO$_3$ substrate. When the virgin SrTiO$_3$ substrate (without proton implantation) is colorless and transparent, the color becomes darker and opaque with the higher implantation dose. This is because the proton implantation produces oxygen vacancies, accompanied by the formation of in-gap states to enhance the absorption of visible lights in the darkened SrTiO$_3$.[22] Although those implanted SrTiO$_3$ substrates contain some oxygen vacancies, they still maintain the insulating nature with resistance $R > 10^8$ Ω. Third, the *a*-LSAT layer was grown on the proton-implanted SrTiO$_3$ substrate by pulsed laser deposition (PLD) under the high-vacuum and room-temperature condition. The high vacuum is required for the formation of oxygen-vacancy-induced quasi-two-dimensional electron system (*q*-2DES) at the amorphous SrTiO$_3$ heterointerface, and the room-temperature deposition is adopted to avoid the high-temperature process that could compromise the ion-implantation effect.[23] Therefore, the proton-implanted *a*-LSAT/SrTiO$_3$ sample is expected to consists of two important charged regions: one is the conventional oxygen-vacancy-induced *q*-2DES close to the heterointerface (red region in **Figure 1(c)**), and the other one is the implanted SrTiO$_3$ layer (green region in **Figure 1(c)**) that is far away from the heterointerface and contains implanted protons with resulted defects.

**Figure 2** summarizes the basic transport properties of *q*-2DES at the proton-implanted *a*-LSAT/SrTiO$_3$ interfaces. To emphasize the modulation of proton-implantation in **Figure 2(a)**, the temperature-dependent sheet resistances, obtained from samples with different proton doses, which are normalized with respect to that of the virgin sample (without proton implantation) as $R_{imp}$ (implanted, *T*)/$R_{vir}$ (virgin, *T*). The nonnormalized temperature-dependent sheet resistances of the virgin and implanted samples are provided in **Figure S1**. The room-temperature sheet resistances (measured at 300 K) are monotonically reduced on increasing proton fluence from 0 to 5E15. However, the low-temperature sheet resistances (measured at 2 K) don't follow this monotonical trend: the low-temperature resistances reach the minimal value when proton fluence is around 1E15. Further increasing the implantation fluence rapidly raises the low-temperature sheet resistances, accompanied by a transition from the metallic behavior (d$R$/d$T$ > 0) to semiconducting (d$R$/d$T$ < 0). Moreover, the sheet resistances are finally out of our measurement range ($R > 10^8$ Ω) when the proton fluence is above 1E16, indicating an insulating behavior. In **Figure 2(b)**, the room-temperature (300 K) and low-temperature (2 K) carrier densities $n_S$ are plotted as a function of proton dose. Our results reveal a clear proton-implantation-induced enhancement on $n_S$, even in the high-fluence samples (up to 5E15) with semiconducting behaviors. Given that the implanted SrTiO$_3$ substrate is not conducting without the on-top *a*-LSAT layer, the observation of enhanced $n_S$ suggests a strong interaction between two charged regions – the proton-implanted SrTiO$_3$ layer and *q*-2DES interface. Also, carrier mobilities $\mu_S$ measured at room temperature and low temperature are compared in **Figure 2(c)**. While the room-temperature $\mu_S$ is almost constant around 3-8 cm$^2$V$^{-1}$s$^{-1}$, the low-temperature $\mu_S$ is very sensitive to the proton fluence. The low-temperature $\mu_S$ reach the maximum value ~ 10,000 cm$^2$V$^{-1}$s$^{-1}$ with the proton fluence around 1E15, corresponding to the minimal low-temperature *R*. Hence, the suppression on metallic behavior in high-dose *a*-LSAT/SrTiO$_3$ heterointerfaces is caused by the reduction on $\mu_S$ rather than $n_S$.

To investigate the location of proton-implanted layer in the SrTiO$_3$ substrate, **Figure 3(a)** presents the simulation results performed by Stopping and Range of Ion in Matter (SRIM).[24] According to the SRIM results, the end of range is at around 300 nm underneath the surface, the proton distribution is a bit deeper than the vacancy region. A detailed SRIM result about vacancy creation is shown in **Figure S2**, which suggests that most of the vacancies are oxygen vacancies.

**Figure 3(b)** compares the $\omega$-$2\theta$ scans of X-ray diffraction (XRD) obtained from samples with different proton fluences. While the (002) peaks (indexed by a dash line) that represent the unaffected part of SrTiO$_3$ is unchanged on increasing the proton dose, the left-side shoulders (indexed by a solid line) resulted from the proton-implanted SrTiO$_3$ layer with defects become significant. Also, those left-side shoulders reveal the lattice expansion of proton-implanted SrTiO$_3$, which can be ascribed to the formation of oxygen vacancies as discussed above.[22,25–30] On the other hand, the cross-section image obtained from transmission electron microscopy (TEM) reveals that the implanted SrTiO$_3$ layer is ~ 450 nm away from the *a*-LSAT/SrTiO$_3$ heterointerface, as shown in **Figure 3(c)** and **Figure S3**. The actual damage depth is deeper than simulation result. This might be because of the channeling effect of proton beam in the crystal lattice. Nevertheless, the fact that proton-implanted SrTiO$_3$ layer is located well below the $q$-2DEG layer is identified. Meanwhile, there are limited number of disorders created at the $a$-LSAT/SrTiO$_3$ interface during the proton implantation to affect the $q$-2DES. It is expected that when the implantation fluence is high enough, the *structural-damage*-induced disorders will raise the energy position of mobility edge with respect to the Fermi level ($E_F$, Fermi energy), leading to Anderson-localization to remove the 2D conductivity at the interface.[31–33] The non-monotonic mobility can be rationalized within an Anderson-localization framework in which extended states exist only for energies above a disorder-dependent mobility edge energy $E_C$.[34–36] By combining Hall densities with the Poisson–Schrödinger Fermi energies and fitting the low-temperature conductivity to $(E_F - E_C)^v$, we find that $E_C$ overtakes $E_F$ near a fluence of 1E15 (see Supplementary Figure S6), coincident with the observed collapse of carrier mobility.

Given the above experimental results, we proposed a model that describes the charge distribution in the proton-implanted $a$-LSAT/SrTiO$_3$ interface as sketched in **Figure 3(d-e)**. When protons are implanted into a bare SrTiO$_3$ substrate, oxygen vacancies (O$_V$) are formed to ionize the positively-charge in-gap states (O$_V^{\bullet\bullet}$) and electrons ($e^-$) at the proton-implanted region. Because of the surface-depletion-induced band bending as shown in **Figure 3(d)**, the thermally excited electrons will be easily trapped by the defect state with O$_V \rightleftharpoons$ O$_V^{\bullet\bullet}$ + 2$e^-$, leading to the insulating property of the proton-implanted SrTiO$_3$ substrate. If the SrTiO$_3$ surface is covered by the $a$-LSAT or $a$-LaAlO$_3$ layer, the surface band will bend in an opposite way to create a potential well for $q$-2DES at the heterointerface as plotted in **Figure 3(e)**. In this case, electrons that are thermally excited from the defect states to conduction band will flow to the heterointerface with O$_V \rightarrow$ O$_V^{\bullet\bullet}$ + 2$e^-$. To substantiate the band-bending model, we performed a self-consistent Poisson–Schrödinger estimate (Supplementary Information). The calculated Fermi energy increases from ~0.05 eV in pristine interfaces to ~0.3 eV at optimal H$^+$ dose, while the characteristic ground-state confinement length remains ~6–10 nm. The higher Fermi energy permits occupation of excited sub-bands, broadening the overall electron distribution and supporting the charge-transfer mechanism proposed in Figure 3(d-e). This is consistent with our observation that both the high-temperature and low-temperature $n_S$ increase on the proton implantation. Hence, two sides of the proton-implantation effect, including *structural damage* and *charge doping*, are presented in the proton-implanted *a*-LSAT/SrTiO$_3$ heterointerface. While *charge doping* plays an important role in low-fluence samples (≤ 1E15), the effect of *structural damage* becomes dominant on increasing the fluence (≥ 1E16).

Given that the ion-implantation-induced *structural damage* with high implantation dose has been well reported,[17–19] we focus on the effect of *charge doping* in low-dose samples. As shown in **Figure 4 (a, b)**, if the bottom implanted SrTiO$_3$ layer acts as the positively-charged donor and the top *q*-2DES as the acceptor with negative charges, the proton-implantation-induced charge doping can be mimicked by the back-gating electrostatic doping, where additional electrons are doped into the top *q*-2DES layer by applying a positive back-gating voltage. In **Figure 4(c)**, the relationship between low-temperature $n_S$ and $\mu_S$ is revealed in the proton-implanted (with fluence no more than 1E15) and back-gated *a*-LSAT/SrTiO$_3$ heterointerfaces. A consistent trend is observed in both cases, where the low temperature $\mu_S$ are improved by increasing $n_S$. One possible explanation is that the increased $n_S$ enhances the screening effect to suppress the disorder-induced scattering. Another possible mechanism is that the positive charges (due to proton implantation) or voltage (from back-gating) underneath the *q*-2DES layer can draw the mobile electrons away from the interfacial defects by Coulomb interaction. Both effects mentioned above may effectively increase carrier mobilities by raising carrier densities. The similar modulation on carrier mobility, mediated by the low-dose proton implantation and positive back-gating voltage, indicates the similar physics of charge doping in both methods.

By modifying the fluence of implanted proton, the carrier mobility of $q$-2DEG at the $a$-LSAT/SrTiO$_3$ interface is improved from 1,000 to 10,000 cm$^2$V$^{-1}$s$^{-1}$. **Figure 5** presents low temperature magneto-transport properties of the selected proton-implanted sample, of which the proton fluence is 1E15 with $n_S$ of $1.12 \times 10^{14}$ cm$^{-2}$ and $\mu_S$ of 8,000 cm$^2$V$^{-1}$s$^{-1}$. When temperature is around 2-3 K and magnetic field $B$ above 6 T, the sample shows Shubnikov-de Haas (SdH) effect featured by the oscillating magnetoresistance in **Figure 5(a)**. If plotting the low-temperature MR as a function of 1/$B$, the oscillating periodicity is around 0.017 T$^{-1}$. The density of high-mobility electron ($n_{SdH}$) that induces the SdH oscillations can be estimated by $n_{SDH} = \frac{2e}{h}\sum f_i$, where $f_i$ frequencies compose the quantum oscillations. Accordingly, $n_{SdH}$ is $\sim 7.5 \times 10^{12}$ cm$^{-2}$ which is much smaller than $n_S$ obtained from Hall measurement. The ratio $n_{SdH}/n_S$ ($\sim 0.1 - 0.3$) falls within the range as reported in the previous works,[37–41] indicating that only the light, high-mobility pockets contribute to the oscillations while the heavier or strongly scattered bands dominate the Hall signal. Such phenomenon with $n_{SdH} < n_S$ is widely observed in the high-mobility $q$-2DES at the SrTiO$_3$-based heterointerface, probably due to the complicated sub-band structure associated with multiple conducting channels. It is also clear to observe that the oscillation longitudinal resistance (ΔR) decreases with increasing temperature as shown in **Figure 5(b)**. The oscillation longitudinal resistance (ΔR) as a function of temperature (**Figure 5(c)**) can be defined as $\Delta R(T) = 4R_0 e^{-\alpha T_D}\alpha T/\sinh(\alpha T)$, where $\alpha = 2\pi^2 k_B/\hbar \omega_C$, $\omega_C = eB/m^*$, $k_B$ is the Boltzmann constant, $\hbar$ is the Planck constant, $\omega_C$ is the cyclotron frequency, $e$ is the elementary charge, $B$ is the magnetic field, $m^*$ is the carrier effective mass, $R_0$ is the non-oscillatory component of $R_S$, and $T_D$ is the Dingle temperature. The fitting of these data by using the equation gives the effective mass $m^* = 0.95 \pm 0.04\ m_e$, where $m_e$ is free electron mass and Dingle temperature $T_D$ = 2.4 ± 0.3 K. This $m^*$ value is consistent with a moderately renormalized $t_{2g}$ band at the $a$-LSAT/STO interface.[37,42–45]

To sum up, we have shown that structural damage and charge doping—two different directions of the ion-implantation effect are both existing. An optimum proton-implanted (1E15 for 50 keV proton) $a$-LSAT/STO sample can lead the high carrier mobility which enables quantum transport oscillations at low temperature. On the other hand, samples with high implant fluences (more than 1E15 protons/cm$^2$) show signs of structural damage, which leads to reduced carrier mobility and insulating behavior. This offers a practical method for adjusting transport properties at SrTiO$_3$-based conducting interfaces in oxide heterostructures, opening avenues for exploring innovative functionalities.

## Materials and Methods

**Sample preparation.** The 0.5 mm thick (001) SrTiO$_3$ (STO) substrate (Crystec) was treated with HF and annealed to obtain defined terrace steps and TiO$_2$-terminated surface. Substrates with proton-implantation were transferred to ion irradiation accelerator prior to the amorphous LSAT (*a*-LSAT) deposition. The pulsed laser deposition method was used for sample preparation. *a*-LSAT was grown at room temperature and high vacuum (10$^{-6}$ Torr). During the growth, a nanosecond KrF 248 nm laser was used with a fluence of 2.0 J cm$^{-2}$ and a repetition rate of 2 Hz.

**Ion irradiation.** A Singletron$^{TM}$ accelerator was used to generate H$_2^+$ ion beams from a hydrogen source bottle and 100kV terminal voltage. 100keV H$_2^+$ was selected by controlling a 90-degree magnetic field. The beam was focused with a quadrupole lens set to a spot size about 50μm × 50μm, and scanned over the whole sample.[46] The irradiation fluence was controlled by the beam current and irradiation dewell time at each pixel.

**Electrical measurements.** Sheet resistance, carrier densities, and carrier mobility were determined using the Van der Pauw method on a physical property measurement system (Quantum Design), which allowed for precise characterization of the electrical properties of the samples. Magneto-transport measurements were conducted over a broad magnetic field range, up to 9 Tesla, to assess quantum oscillations phenomenon on the transport measurements.

## Acknowledgements

This research is supported by the Science and Engineering Research Council of A*STAR (Agency for Science, Technology and Research) Singapore, under Grant (M22L1b0110). The authors acknowledge financial support from the Singapore Ministry of Education (MOE) Academic Research Fund Tier 3 Grant (MOE-MOET32023-0003) "Quantum Geometric Advantage" and the Tier 2 Grants (MOE-T2EP50123-0013 and MOE-T2EP50221-0009).

## Data availability

The data that support the findings of this study are available from the corresponding authors upon request.

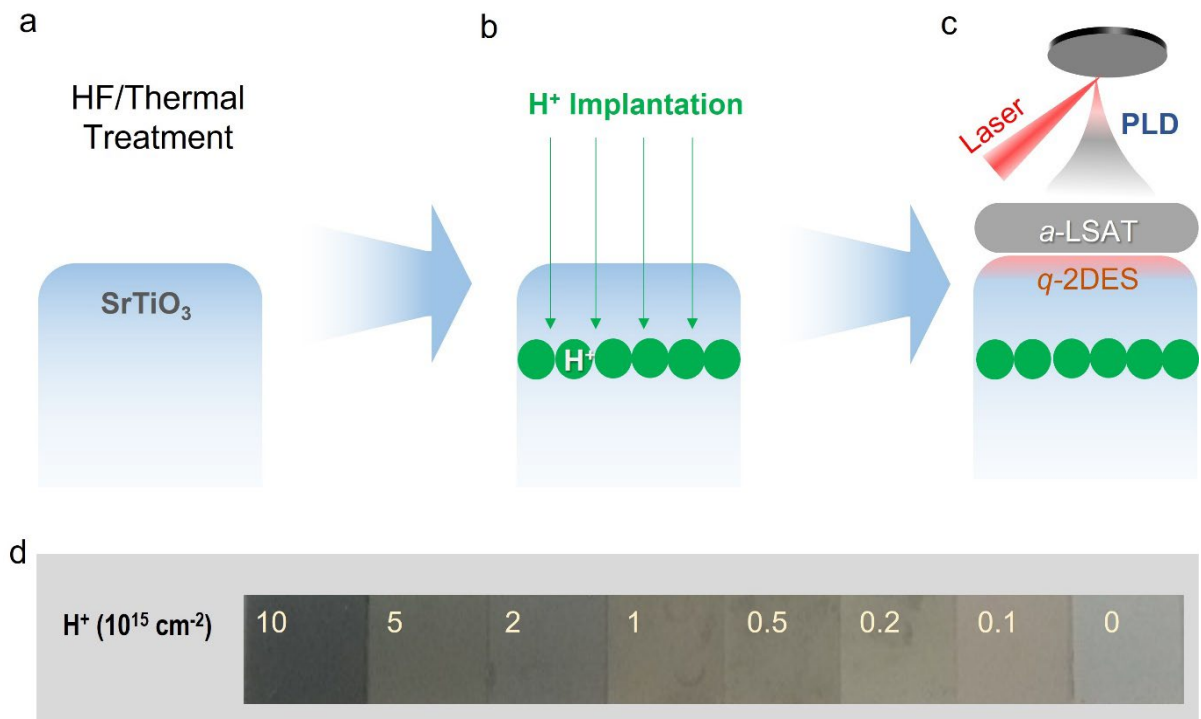

**Figure 1. Schematic of the sample preparation and optical images of the samples before and after proton implantation**. (a) Buffered HF and thermal annealing treatment with the STO wafer. (b) Schematic of proton implantation in the treated STO. (c) PLD of a-LSAT process on the treated and implanted STO wafer. (d) Optical images of the STO before (right most, marked as 0) and after proton implantation with different doses from 1E14 to 1E16 (from right to left).

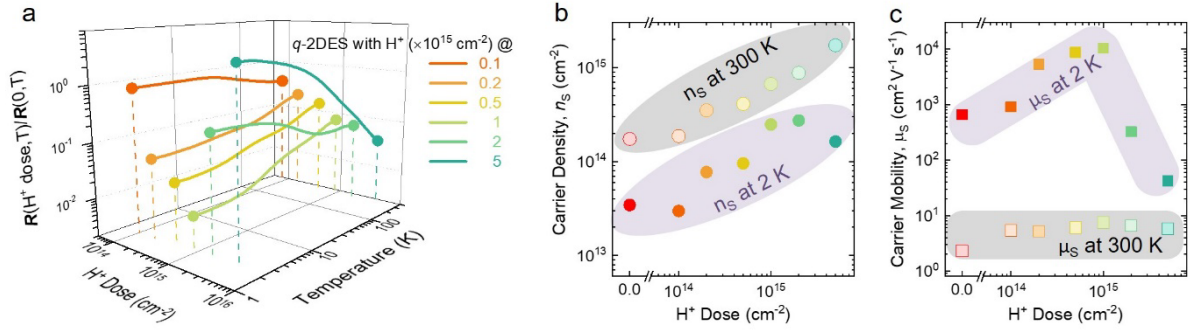

**Figure 2. Transport properties of the *a*-LSAT/SrTiO3 interfaces.** (a) Relative resistances the proton implanted samples comparing with the virgin sample. (b) Carrier densities and (c) mobilities of the virgin and implanted samples at 2K and 300K.

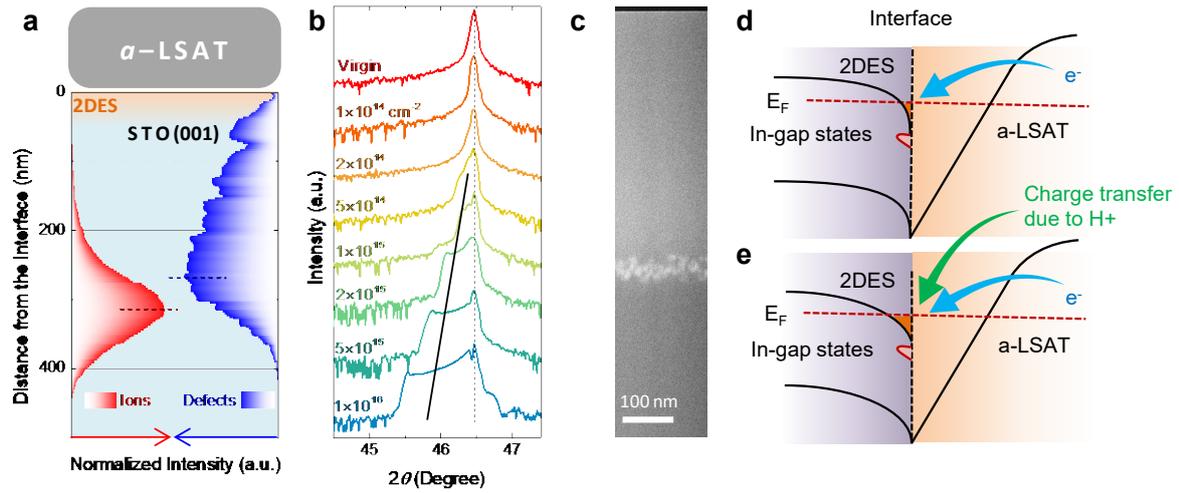

**Figure 3. Ion and vacancy distribution of the proton implanted STO wafer.** (a) SRIM result of 50 keV proton in STO. (b) XRD of the STO before and after proton implantation. (c) TEM image of the cross section of the irradiated STO. (d) Band bending model of *a*-LSAT/STO with in-gap states with low carrier density. (e) and with higher carrier density due to H+ charge transfer.

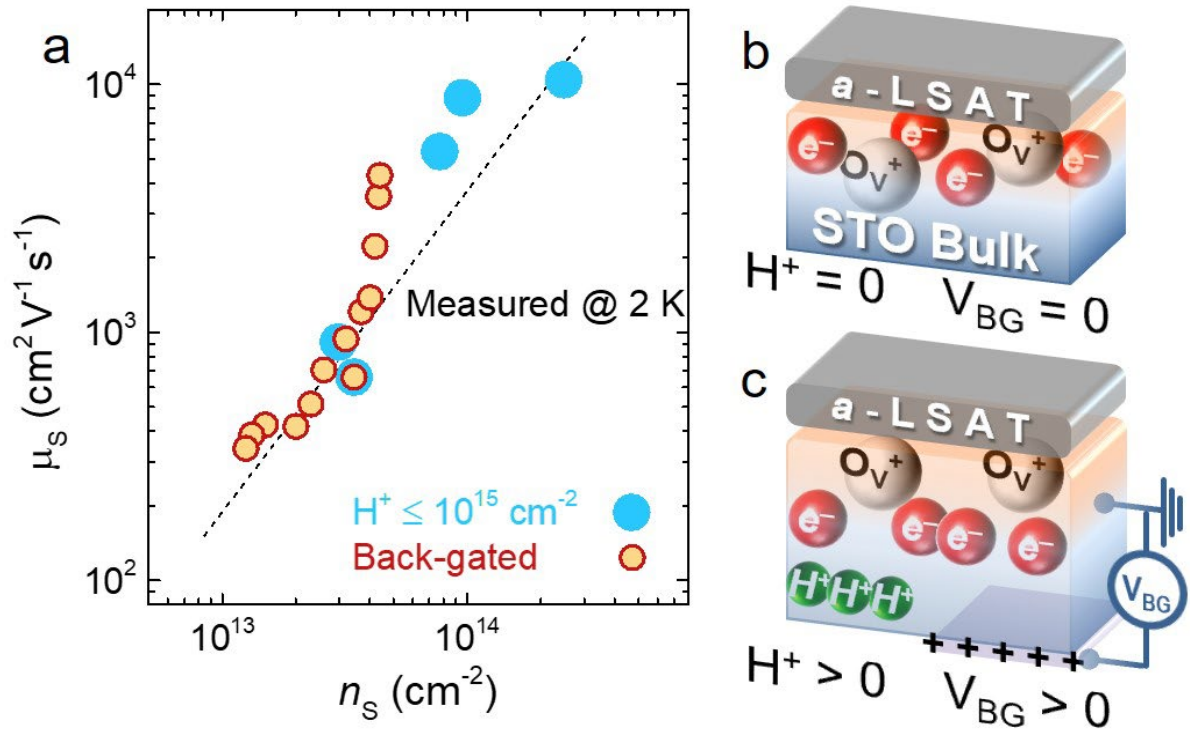

**Figure 4. Comparison of the proton implanted and back-gated induced transport**. (a) Relationship between $\mu_S$ and $n_S$ at 2 K for the ion implanted with back-gated induced transport. Schematics to show the difference in STO electron carrier density in (b) virgin and (c) implanted samples with back-gate.

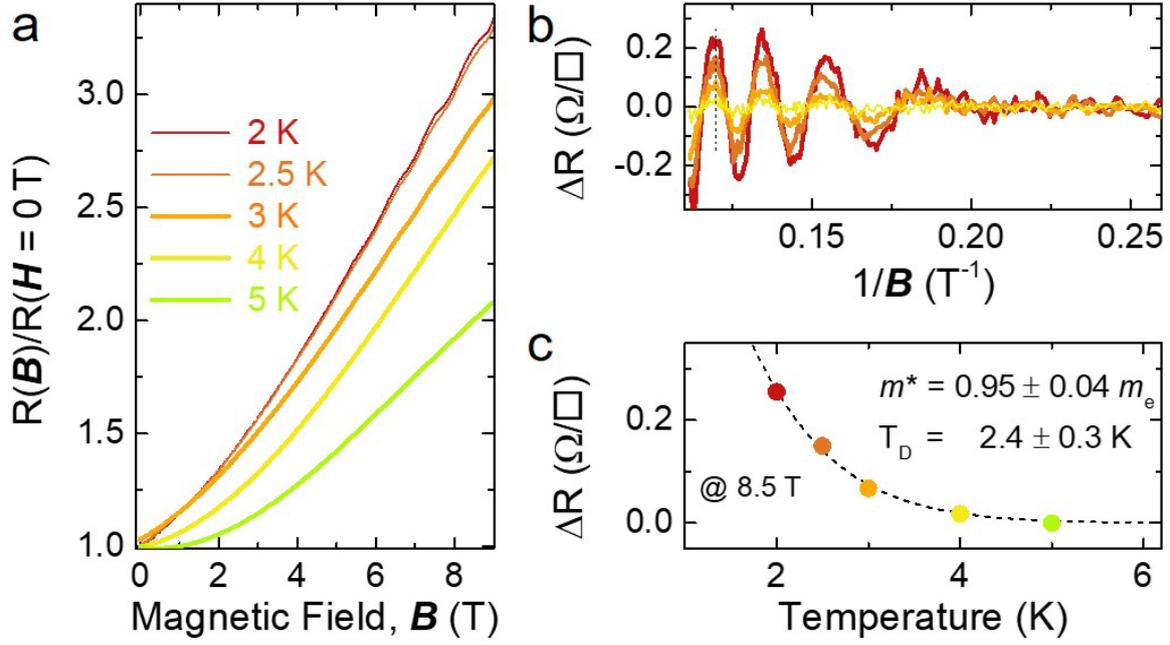

**Figure 5. Low temperature magneto-transport properties.** (a) Longitudinal resistance of the selected proton-implanted sample as a function of the field for different temperature ranging from 2 K to 5 K. (b) Inverse-field dependence of the oscillating longitudinal resistance (ΔR). (c) Temperature dependence of the longitudinal resistance (ΔR) for 8.5 T magnetic field. Symbols are the experimental data, and the solid lines are the Lifshitz-Kosevich (L-K) fit. Note: Figures 5(b) and 5(c) follow the same colour scheme as indicated in the legend of Figure 5(a). Light green colour plot (5K) in Figure 5 (a) doesn't have any oscillations, therefore it is not shown in the Figure 5 (b).

# Supplemental Information

# Modulation of Quantum Transport in Complex Oxide Heterostructures with Proton Implantation

Haidong Liang[1,2,#], Ganesh Ji Omar[1,#], Kun Han[3], Andrew A. Bettiol[1,2,*], Zhen Huang[3,4,*], Ariando[1,*]

[1]Department of Physics, National University of Singapore, Singapore 117551, Singapore

[2]Centre for Ion Beam Applications (CIBA), Department of Physics, National University of Singapore, Singapore 117542, Singapore

[3]Leibniz International Joint Research Center of Materials Sciences of Anhui Province, Institutes of Physical Science and Information Technology, Anhui University, Hefei 230601, China

[4]Stony Brook Institute at Anhui University, Anhui University, Hefei 230039, China

[#] These authors contributed equally to this work.

[*] Contact Author: a.bettiol@nus.edu.sg, huangz@ahu.edu.cn, phyarian@nus.edu.sg

**Control experiments on different implantation & deposition sequences:**

To clarify the role of implantation relative to the *a*-LSAT deposition step, we systematically examined five different cases, as summarized in **Figure S1**. Bare SrTiO₃ without implantation remains highly insulating, with the sheet resistance above the instrumental measurement limit of about 5–6 MΩ. Proton implantation of SrTiO₃ alone also leaves the substrate insulating, confirming that implantation does not by itself generate conduction. When an *a*-LSAT layer is deposited without implantation, a conventional *q*-2DES forms at the interface, giving a sheet resistance of about 15 *k*Ω at room temperature, consistent with prior reports.

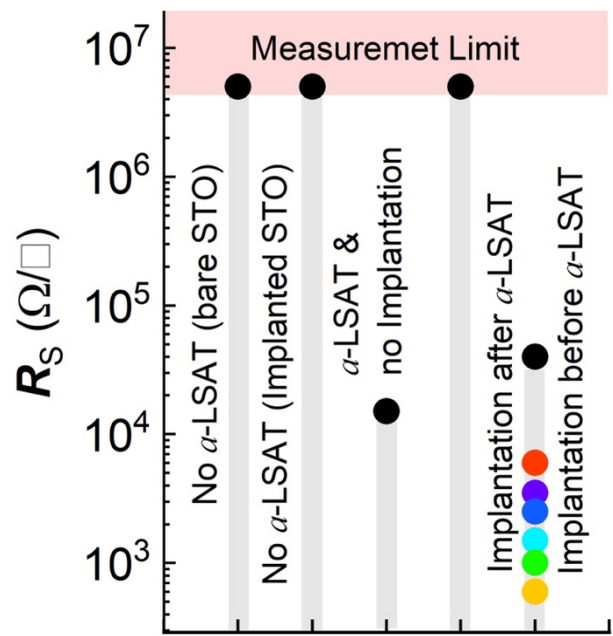

**Figure S1. Role of Implantation sequence.** Sheet resistance ($R_S$) measured for five different scenarios: (i) bare SrTiO₃ without implantation, (ii) proton-implanted SrTiO₃ without *a*-LSAT, (iii) *a*-LSAT/STO without implantation, (iv) *a*-LSAT/STO with implantation performed after deposition, and (v) *a*-LSAT/STO with implantation performed before deposition. Cases (i), (ii), and (iv) remain insulating with $R_S$ above the measurement limit (~5–6 MΩ). Case (iii) exhibits the conventional *q*-2DES with $R_S \approx 15$ kΩ, while case (v) yields tunable transport properties as studied in the main text.

If implantation is performed after *a*-LSAT deposition, the initially formed *q*-2DES is destroyed as implanted protons and associated defects traverse the interfacial region, rendering the system insulating. In contrast, when implantation is carried out before deposition, tunable transport properties emerge from the competition between charge doping and disorder, as described in the main text. All measurements were performed using the van der Pauw method in a PPMS system. The insulating cases show resistances above the detection limit, while only the deposited or pre-implanted heterostructures exhibit measurable conduction. Among these, the pre-implanted case provides the systematic modulation of transport properties discussed in the manuscript. Figure S2 shows the absolute temperature-dependent sheet resistances of the virgin and proton implanted samples.

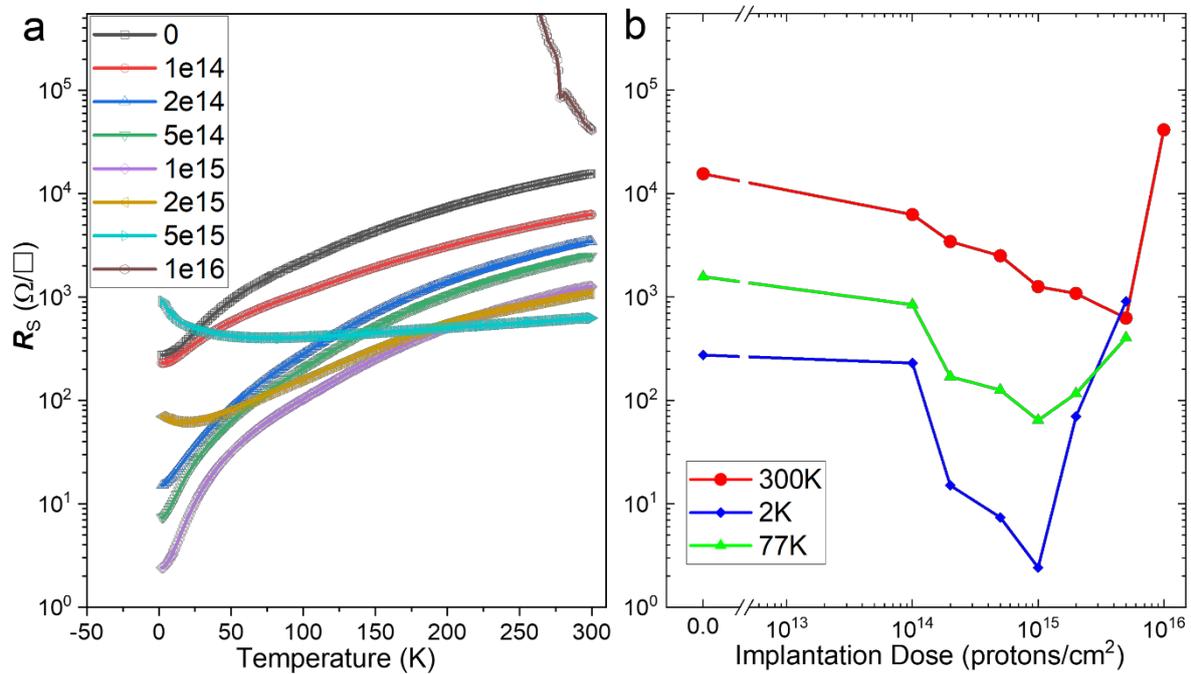

**Figure S2. Transport properties of the *a*-LSAT/SrTiO3 interfaces.** a. Temperature-dependent sheet resistances ($R_S$) of the virgin (black) and proton implanted samples. b. Sheet resistances of samples with different ion implantation fluences at 2K (blue), 77K (green) and 300K (red).

Figure S3 presents the SRIM simulation of defect distributions, revealing that the majority of the defects generated during proton implantation are oxygen vacancies.

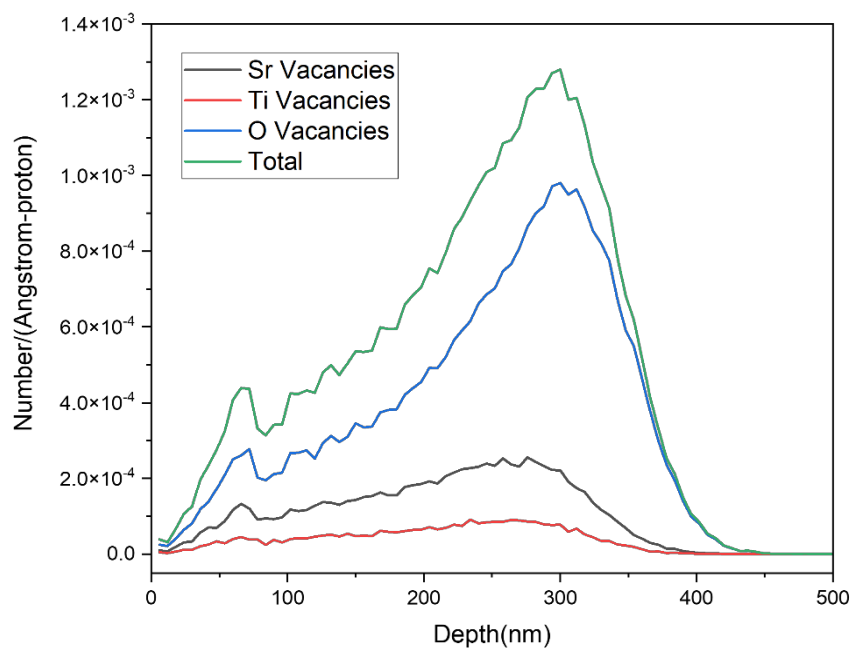

**Figure S3.** SRIM simulation illustrating the detailed defect distributions of various components within the SrTiO$_3$ (STO) structure, providing insights into the spatial profiles and concentrations of defects resulting from proton implantation.

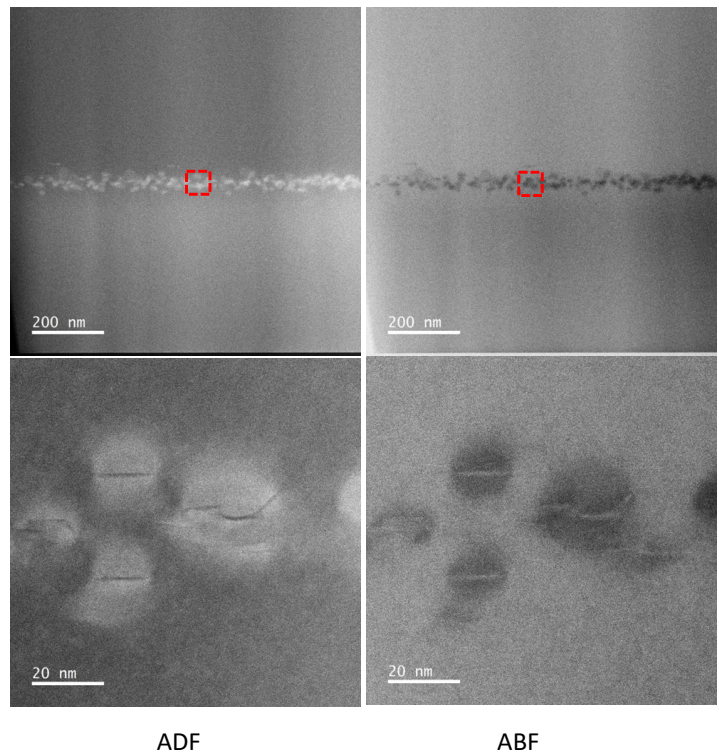

| ADF | ABF |

**Figure S4.** Additional TEM images showing the cross-section of the proton-implanted SrTiO₃ (STO) substrate at larger scales. These images provide a broader view of the structural features, highlighting the distribution of defects and any microstructural variations induced by proton implantation.

**Framework for Quantifying Confinement and Carrier Redistribution**

We performed one-dimensional self-consistent Poisson- Schrödinger calculation to estimate the confining potential and the spatial extent of the electron density for the pristine and proton-implanted $a$-LSAT/STO interfaces.[1,2]

1. **Pristine $a$-LSAT/STO interface (reference case)**

At the pristine LSAT/STO interface, the confinement arises mainly from oxygen vacancy-induced $q$-2DES in our case. The Poisson equation:

$$\frac{d^2V(z)}{dz^2} = -\frac{\rho(z)}{\epsilon_0\epsilon_r}$$

where, $V(z)$ is the electrostatic potential, $\rho(z) = e[n(z) - N_D^+(z)]$ is the net charge density, $\epsilon_r \approx 300$ (at low temperature for STO).

Schrödinger equation (effective mass approximation):

$$-\frac{\hbar^2}{2m^*}\frac{d^2\varphi_i(z)}{dz^2} + eV(z)\varphi_i(z) = E_i\varphi_i(z)$$

with $m^* \sim 0.7 - 1.0\ m_e$ (light $d_{xy}$ band) and larger for $d_{xz/yz}$ ($1.5 - 2.0\ m_e$).

Boundary condition: band bending near interface forms a triangular-like potential well.

Solution: quantized subband ($d_{xy}$ lowest, $d_{xz/yz}$ higher), carrier density $\sim n_{2D}$ confined within ~5–10 nm of interface.

2. **Implanted $a$-LSAT/STO case**

After proton implantation, additional positive defect charges (mainly O vacancies) are introduced at depth $z_d \sim 300 - 400$ nm. This modifies the Poisson equation:

$$\frac{d^2V(z)}{dz^2} = -\frac{\rho(z)}{\epsilon_0\epsilon_r}[n(z) - N_D^+(z) - N_{imp}^+\ \delta(z - z_d)]$$

here, $N_{imp}^+$ is the implanted donor density. This remote positive sheet acts analogously to a back gate, pulling the conduction band downward across hundreds of nm. In the Poisson solution, this produces a deeper and wider triangular potential well at the interface. In the Schrödinger solution, more subbands can be occupied, particularly $d_{xz/yz}$ orbitals with larger spatial extent. For a triangular potential well, the characteristic confinement width is given by:

$$L_c \approx \left(\frac{\hbar^2\epsilon_0\epsilon_r}{2m^*e^2n_s}\right)^{1/3}$$

And the 2D Fermi energy is $E_F = \pi \hbar^2 n_s / m^*$. In the pristine a-LSAT/STO with $n_s \sim 1 \times 10^{13}$ cm$^{-2}$, $m^* \sim 0.7 m_e$, $\epsilon_r \sim 300$. We get $L_c \sim 5 - 10$ nm. In the implanted case, $n_s$ increases nearly an order of magnitude $1 \times 10^{14}$ cm$^{-2}$). Plugging that into the formula:

$$L_c^{imp} \sim \left( \frac{n_s^{pristine}}{n_s^{imp}} \right)^{1/3} L_c^{pristine}$$

Thus, the potential well becomes deeper and wider in the implanted case and the spatial extent of the *q*-2DES increases significantly, which explains the observed higher mobility (screening of disorder by larger density and more extended wavefunctions) as shown in **Figure S5**.

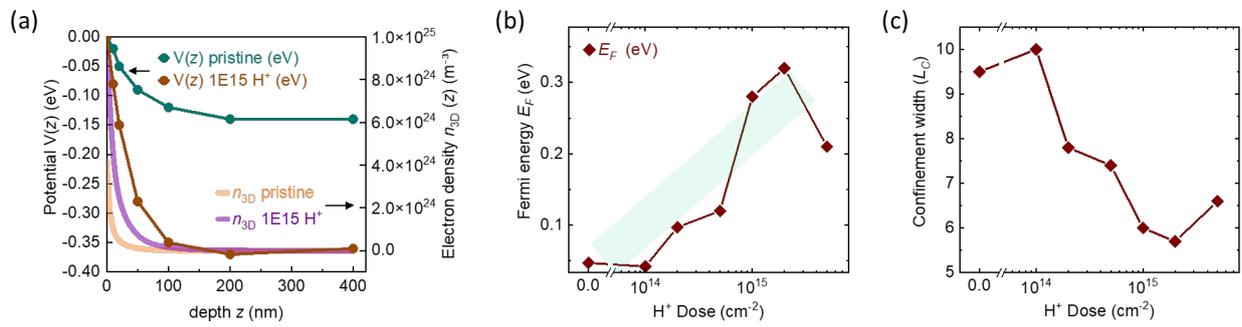

**Figure S5.** (a) The plot shows potential $V(z)$ (left axis) and 3D carrier density (z) (right axis) versus depth in STO substrate. The implanted sample exhibits a deeper and wider potential well and a broader electron distribution, (b) Fermi energy ($E_F$) rises up with increasing H$^+$ implantation, (c) characteristic confinement length of the ground sub-band decreases modestly, while the higher-subband occupation accounts for the broadened total carrier profile.

**Mobility-Edge Analysis of the Disorder–Doping Crossover**

To evaluate the competition between disorder and charge doping, we extracted the Fermi energy $E_F$ from the sheet carrier density using $E_F = \pi\hbar^2 n_s/m^*$, where $m^* = 0.95 m_e$ (from SdH analysis). The mobility-edge energy $E_C$ was obtained by fitting the low temperature sheet conductivity to the scaling form $\sigma = \sigma_0(E_F - E_C)^v$, with critical exponent $v = 1.0 \pm 0.1$. The fitted $E_C$ increases nearly linearly with proton fluence, reflecting the rising defect density, while $E_F$ rises due to donor doping. The intersection of these two energy scales at a fluence near 1E15 signals the transition from extended to localized states.

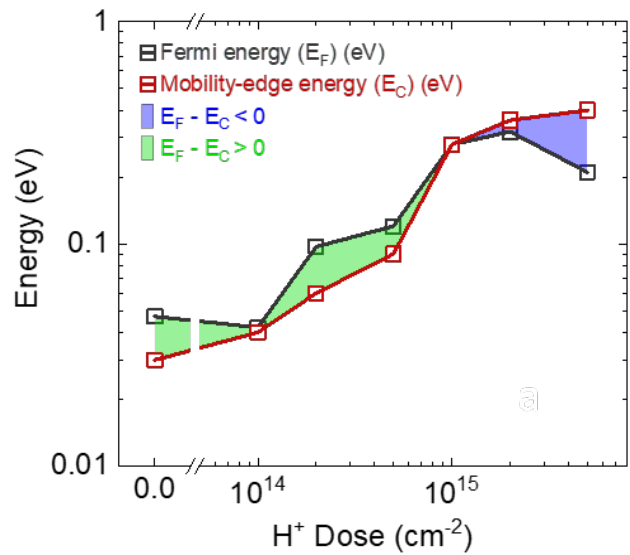

**Figure S6.** Fermi energy ($E_F$) and Mobility-edge energy ($E_C$) is plotted against H+ implantation dose and the intersection near $1 \times 10^{15}$ cm$^{-2}$ marks the crossover transition where disorder (rising $E_C$) over takes doping (rising $E_F$), producing the observed drop in mobility.